# Molecular Photonics Meets OLED Technology: The Untapped Potential of Lanthanide Emitters

Arseny Yu. Gladkikh[a], Aleksandr Yu. Frolov[a], Valentina V. Utochnikova[a]
[a] Lomonosov Moscow State University, Leninskie Gory 1, Moscow, Russia, 199991

## Abstract

Lanthanide-based emitters are attracting increasing interest for their potential in high-performance OLEDs due to their narrow emission bands, high photostability, and functional versatility. However, their low radiative rates and poor charge transport properties remain key challenges for efficient device integration. Here, we report the design, synthesis, and characterization of dual-emissive $Eu^{3+}$–$Yb^{3+}$ coordination compounds that exhibit temperature-dependent luminescence intensity ratios (LIR) between the visible and near-infrared regions. These materials enable simultaneous optical thermometry and electroluminescence, with LIR values showing excellent agreement with external thermocouple measurements over a broad temperature range. OLED devices incorporating these complexes demonstrate stable operation, and their luminescence properties are further enhanced by integration with engineered plasmonic metasurfaces. This integration leads to a significant increase in emission directionality and a reduction of the excited-state lifetime via Purcell enhancement, without compromising spectral integrity. Our findings highlight the promise of lanthanide-based coordination compounds as multifunctional emitters for next-generation optoelectronic devices and suggest new pathways toward intelligent OLED architectures with integrated sensing and directional control.

## Introduction

Organic light-emitting diodes (OLEDs) have evolved into a mature display and lighting technology, driven by advances in molecular design, device engineering, and charge-transport materials. Most current high-efficiency or thermally activated delayed fluorescence (TADF) materials to achieve near-unity internal quantum efficiencies. However, these systems face intrinsic limitations related to broad emission spectra, stability, and emission wavelength tunability—particularly in the near-infrared (NIR) region.

Lanthanide-based coordination compounds offer a fundamentally different approach to light emission. Their sharp line-like emission bands, high color purity, and intrinsic NIR transitions (e.g., in $Yb^{3+}$, $Nd^{3+}$, $Er^{3+}$) are direct consequences of 4f–4f electronic transitions shielded from the ligand environment. Moreover, lanthanide complexes exhibit remarkable photostability and can be engineered to provide dual emission or temperature-dependent luminescence intensity ratios (LIR), enabling functionalities such as ratiometric thermometry and optical encoding. Despite these advantages, their integration into OLED devices remains remarkably underdeveloped.

Several factors have hindered the widespread adoption of lanthanide complexes in OLEDs. Most notably, the intrinsically low radiative transition probabilities of 4f–4f transitions result in long excited-state lifetimes (typically in the microsecond range) and low electroluminescence efficiencies. These characteristics make exciton quenching, charge imbalance, and energy back-transfer to host materials particularly problematic. Furthermore, efficient sensitization of lanthanides under electrical excitation is still not fully understood and requires precise control of molecular energetics and device architecture.

In parallel, the field of nanophotonics has provided powerful tools to tailor light–matter interactions via optical confinement, field enhancement, and photonic environment engineering. In particular, plasmonic and dielectric metasurfaces have demonstrated the ability to enhance radiative decay rates (Purcell effect), improve outcoupling efficiency, and introduce emission directionality in organic emitters. These strategies have been successfully applied to OLEDs based on conventional phosphorescent and TADF emitters. Moreover, numerous studies have demonstrated enhanced photoluminescence and shortened excited-state lifetimes of lanthanide complexes in the presence of metallic nanostructures.

Yet, the application of such photonic and plasmonic strategies to lanthanide-based OLEDs is virtually nonexistent. To the best of our knowledge, only a single study has explored the integration of a lanthanide emitter with a structured metallic surface in an electroluminescent device. This striking gap highlights both the difficulty and the unexplored potential of combining lanthanide photophysics with nanoscale light control under electrical excitation.

In this review, we examine the opportunities and challenges of developing lanthanide-based OLEDs through the lens of modern photonic and device engineering. We first analyze the fundamental photophysical properties of lanthanide complexes and their implications for electroluminescent performance. We then discuss existing OLED architectures and design strategies for accommodating lanthanide emitters, as well as the prospects for applying nanophotonic techniques—including metasurface integration—to overcome current limitations. Through this synthesis, we aim to articulate a roadmap for realizing high-performance lanthanide-based OLEDs that go beyond current paradigms, enabling multifunctional light sources with built-in sensing, directionality, and spectral selectivity.

### Section 1. Preliminary Information

#### 1.1. Fundamentals of Lanthanide Photophysics

Lanthanide ions exhibit exceptional photophysical behavior due to their ability to undergo 4f-4f electronic transitions. The electron configuration of lanthanides ([Xe]4f$^n$, n = 0–14) results in a remarkably large number of closely spaced excited states – up to ~3000 microstates for ions such as $Eu^{3+}$ and $Tb^{3+}$ – as illustrated in the Dieke diagram (**Figure 1**)[1].

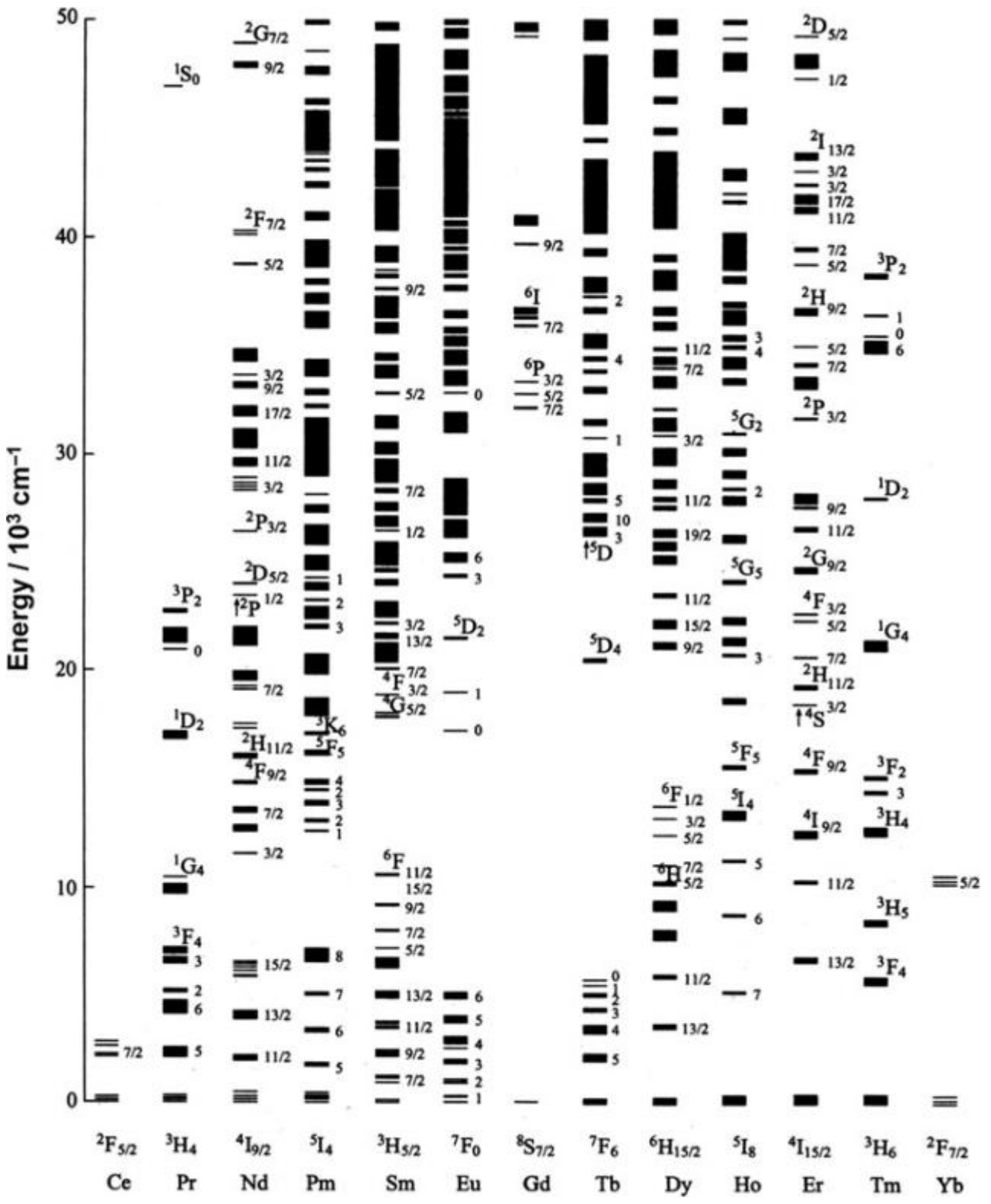

**Figure 1.** Dieke Diagram.

A hallmark of lanthanide photoluminescence is the spectral invariance of emission bands with respect to the chemical environment. This is a direct consequence of the shielding of the 4f orbitals by the filled $5s^2$ and $5p^6$ subshells. However, f-f transitions are formally Laporte-forbidden, leading to exceptionally long excited-state lifetimes on the order of $10^{-6}$ to $10^{-2}$ seconds[2].

To overcome the intrinsically low absorption cross-sections of lanthanide ions, coordination with organic ligands is employed to enable efficient light harvesting via the so-called antenna effect[3]. In this mechanism, illustrated in the Jablonski diagram (**Figure 2a**), the organic ligand absorbs light and undergoes rapid internal conversion to its lowest singlet excited state ($S_1$). From there, intersystem crossing populates the triplet state ($T_1$), which can efficiently transfer its energy to the resonance level of the lanthanide ion. This results in characteristic sharp-line emission from the lanthanide center[4].

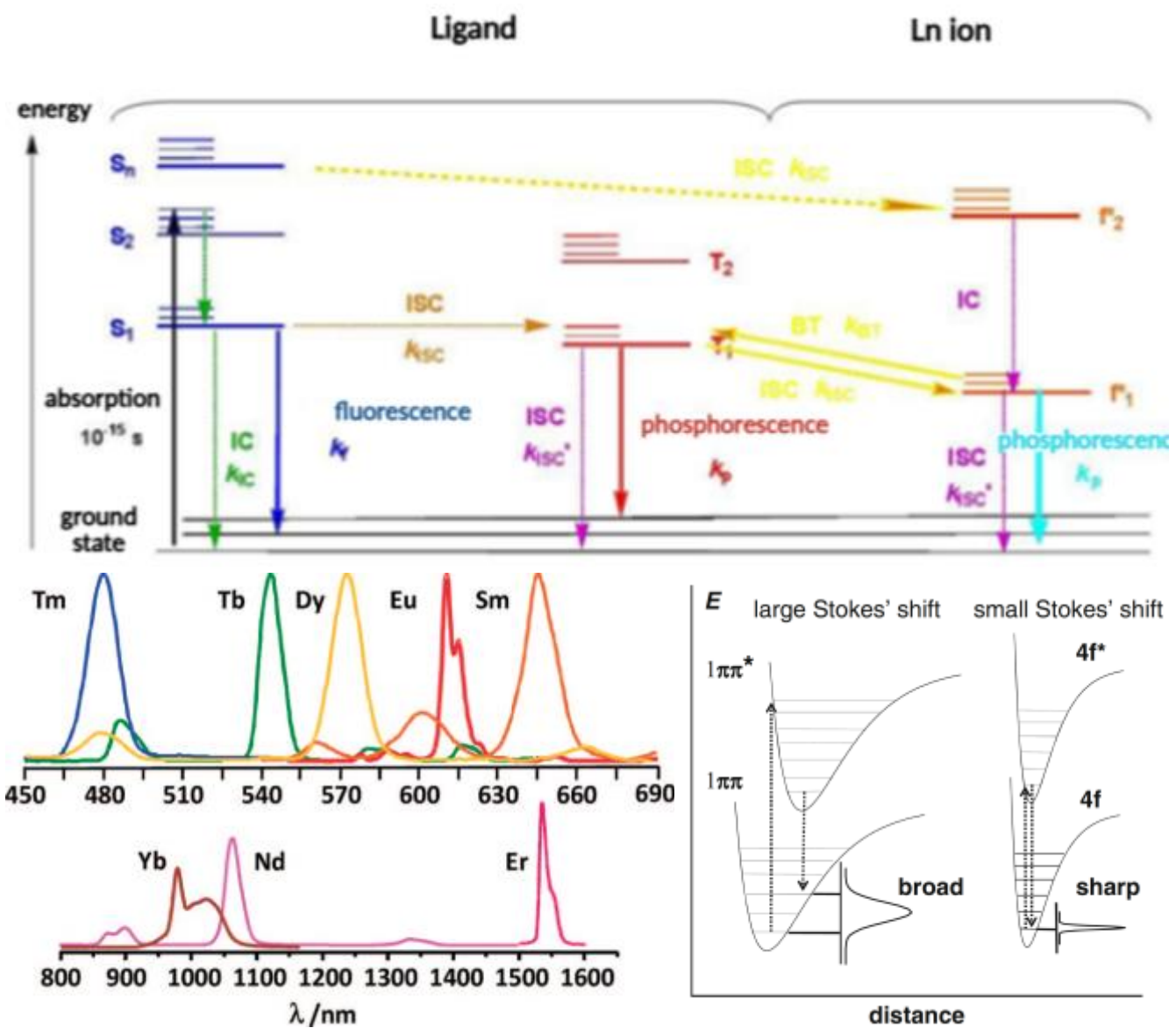

**Figure 2.** a) Jablonski diagram. b) The luminescence spectra of lanthanide ions. c) The dependence of the energy of the organic molecule (left) and the lanthanide complex (right) on the nuclear coordinates in the ground and excited state.

Energy transfer from the ligand triplet state to the lanthanide ion can proceed via two primary mechanisms: Dexter electron exchange, which requires orbital overlap and is favored when the ligand is covalently or coordinatively bound to the metal; and Förster resonance energy transfer (FRET)[5], which operates through long-range dipole–dipole interactions and is highly sensitive to ligand structure, spatial arrangement, and environmental factors[6]. Compared to traditional organic luminophores, lanthanide complexes show minimal structural relaxation in the excited state (**Figure 2a**)[7], resulting in remarkably narrow emission bands (typically <10 nm FWHM) that are largely unaffected by the ligand field (**Figure 2b**)[8]. This makes lanthanide complexes ideal candidates for applications requiring spectral purity and environmental robustness.

The key performance parameter in lanthanide luminescence is the photoluminescence quantum yield (PLQY), defined as the ratio of emitted to absorbed photons. For coordination complexes, PLQY is determined by both the sensitization efficiency (energy transfer from ligand to metal) and the intrinsic quantum yield of the lanthanide ion, as expressed by:

$$\boldsymbol{PLQY = \eta_{sens} \times Q_{Ln}^{Ln}} \quad (\mathbf{1})$$

The intrinsic quantum yield reflects the balance between radiative and non-radiative decay processes[1]:

$$\boldsymbol{Q_{Ln}^{Ln} = \frac{k_{rad}}{k_{rad} + \sum k_{nr}} = \frac{\tau_{obs}}{\tau_{rad}}} \quad (\mathbf{2})$$

Here, $k_{rad}$ is the radiative decay rate, and $k_{obs}$ is the observed decay rate, which includes all non-radiative contributions. The sensitization efficiency is defined as:

$$\eta_{sens} = QY^{L}_{Ln} / QY^{Ln}_{Ln} \quad (3)$$

Efficient energy transfer from ligand to lanthanide requires a suitable triplet energy level on the ligand. An optimal energy gap between the ligand triplet and the lanthanide excited state is typically 2000–4500 $cm^{-1}$ [9].

The excited-state lifetime ($\tau_{obs}$) is another crucial parameter, representing the average residence time of the ion in the excited state:

$$\tau_{obs} = \frac{1}{k_{rad} + \sum k_{nr}} \quad (4)$$

Typical lifetimes for fluorescence and phosphorescence are $10^{-10}$–$10^{-8}$ s and $10^{-6}$–$10^{1}$ s, respectively[5]. Lanthanide complexes exhibit excited-state lifetimes in the range of 0.1–1 ms[10]. These long lifetimes arise from the forbidden nature of f–f transitions and, while they are advantageous for time-resolved spectroscopy and sensing, they also limit the brightness and current efficiency in electroluminescent devices. Thus, enhancing $k_{rad}$ and minimizing $k_{nr}$ are critical strategies in designing high-performance lanthanide emitters.

### 1.2. OLED Structure and Operating Principles

Organic light-emitting diodes (OLEDs) are electroluminescent devices that convert electrical energy into light through the radiative recombination of electrons and holes injected into the emissive layer (EML). Due to their tunable optical properties, solution processability, and potential for flexible displays, OLEDs represent one of the most advanced platforms in modern optoelectronics.

A typical OLED comprises a multilayer heterostructure (Figure 4), in which organic functional layers are sequentially deposited between a transparent anode and a reflective cathode[11]. The general architecture includes, in order: a substrate (commonly glass), anode, hole injection layer (HIL), hole transport layer (HTL), emissive layer (EML), electron transport layer (ETL), electron injection layer (EIL), and corresponding blocking layers (HBL and EBL).

The anode, typically made of indium tin oxide (ITO), must exhibit high optical transparency to maximize light outcoupling. It injects holes into the HOMO level of the adjacent organic layer under applied forward bias. The cathode, usually composed of aluminum, injects electrons into the LUMO of the EIL and ETL. To reduce injection barriers and improve energy level alignment, interfacial modifiers such as LiF are often introduced at the cathode interface.

Charge carriers are transported across the respective transport layers to the EML, where recombination occurs. Blocking layers serve to confine charge carriers and excitons within the EML, thus enhancing recombination efficiency. Often, a single material can serve dual functions, e.g., ETL and HBL, provided it exhibits suitable electron mobility and hole-blocking characteristics. The emissive layer is the core functional unit of the device, where exciton formation and subsequent radiative decay take place[12].

The emission process in OLEDs proceeds through a cascade of steps: (i) charge injection from the electrodes, (ii) carrier transport through the organic layers, (iii) electron–hole recombination and exciton formation in the EML, and (iv) radiative relaxation of excitons (**Figure 5**)[13].

Under applied bias, electrons are injected from the cathode and transported through the LUMO levels of the EIL and ETL, while holes are injected from the anode and move through the HOMO levels of the HIL and HTL. For efficient charge injection and minimal energy loss, the energy offsets between electrode Fermi levels and the adjacent molecular orbitals should not exceed 0.5 eV[14].

Exciton formation occurs when an electron and a hole localize on the same or adjacent molecules in the EML. This process leads to the creation of a Coulomb-bound electron–hole pair, which can exist in either a singlet ($S_1$) or triplet ($T_1$) excited state, depending on the relative spin orientation. According to spin statistics, 25% of the excitons formed are singlets and 75% are triplets (Figure 6). Efficient utilization of both singlet and triplet excitons is essential for maximizing OLED efficiency. Therefore, materials capable of harvesting triplet excitons via phosphorescence or thermally activated delayed fluorescence (TADF) are critical to device design[15].

Key performance indicators for OLED devices include the electroluminescence (EL) spectrum, power efficiency (PE), current efficiency (CE), and external quantum efficiency (EQE). EL spectra provide insight into the emission color and recombination dynamics within the EML.

- Power efficiency (PE) is defined as the luminous flux emitted per unit electrical power consumed and is expressed in lumens per watt ($lm \cdot W^{-1}$).
- Current efficiency (CE) denotes the luminous intensity per unit current density ($cd \cdot A^{-1}$).
- External quantum efficiency (EQE), the most comprehensive efficiency parameter, reflects the ratio of emitted photons to injected electrons:

$$EQE = r \cdot \eta_r \cdot \Phi_{PL} \cdot \eta_{out} \quad (5)$$

Here, $r$ represents the charge balance factor, $\eta_r$ is the exciton formation efficiency (typically 100%), $\Phi_{PL}$ is the photoluminescence quantum yield of the EML, and $\eta_{out}$ is the light outcoupling efficiency, which is generally limited to ~30% in conventional planar OLEDs.

Device electrical characteristics are typically evaluated via current–voltage (J–V) and luminance–voltage (L–V) curves. The turn-on voltage is commonly defined as the voltage at which luminance reaches 1 $cd \cdot m^{-2}$ or current density exceeds 0.1 $mA \cdot cm^{-2}$ [16].

Together, these parameters provide a comprehensive picture of the electronic, optical, and energy conversion efficiencies of OLED architectures. Optimization of energy-level alignment, carrier balance, and radiative pathways remains central to advancing next-generation OLED emitters.

### 1.3. Lanthanide Coordination Compounds in OLEDs

Lanthanide coordination compounds represent a promising class of emitters for organic light-emitting diodes (OLEDs). The first OLED based on lanthanide complexes was reported by J. Kido's group in 1990[17]. Since then, these materials have attracted growing attention due to several intrinsic advantages over traditional fluorescent, phosphorescent, and TADF emitters.

One of the key features of lanthanide-based emitters is their ultra-narrow emission bands, with full width at half maximum (FWHM) typically below 10 nm. This is a direct consequence of the shielded 4f orbitals, which are only weakly perturbed by the ligand field[18], and results in rigidly defined excited-state geometries. In contrast, conventional emitters often display FWHM values of 50–100 nm. Such narrow emissions are particularly advantageous for realizing high color purity in display applications.

Another critical benefit lies in the invariant nature of the emission wavelengths across different ligand environments. The emission maxima of $Eu^{3+}$ and $Tb^{3+}$ complexes, for instance, are located at 612 nm and 545 nm, respectively—perfectly aligned with the red and green primaries required for RGB OLED applications[19]. Moreover, photoluminescence quantum yields (PLQY) approaching 100% have been achieved for these complexes[20,21], underscoring their potential for efficient light emission.

A major challenge in using lanthanide complexes in OLEDs lies in charge carrier balance within the emissive layer. Since exciton formation in these materials generally occurs on the organic ligand framework, efficient device performance requires careful tuning of both electron and hole transport properties. This is often addressed via ligand design, incorporating electron-donating or -withdrawing functional groups to enhance carrier mobility. For instance, our group has employed carbonyl[22], tetrazolate[23], and benzoate[24] anionic ligands, in combination with phenanthroline-based neutral ligands[23], to improve charge balance and enhance device performance.

Despite these advances, lanthanide-based devices still underperform compared to Ir(III) phosphorescent OLEDs in terms of brightness and efficiency. One of the limiting factors is the intrinsically long excited-state lifetimes of lanthanide emitters, typically in the range of 0.1–1 ms—2–3 orders of magnitude longer than those of Ir-based phosphors (~1 μs)[10]. These long lifetimes reduce the radiative recombination rate of charge carriers, thereby increasing the probability of detrimental processes such as triplet–triplet annihilation and device degradation[25].

To address this, strategies are being developed to reduce the excited-state lifetime (τ), which according to **Equation (4)** can be achieved by increasing the radiative decay rate ($k_{rad}$) or the nonradiative decay rate ($k_{nr}$). While introducing nonradiative pathways such as vibrational quenching or energy transfer to adjacent levels can shorten τ, this approach often compromises PLQY[26]. A more promising strategy involves partially lifting the parity-forbidden nature of f–f transitions. This can be achieved by symmetry-breaking in the coordination environment or via mixing of f and d orbitals, rendering f–f transitions partially allowed[27]. Another emerging method is the application of plasmonic enhancement, which has been shown to reduce lifetime and improve electroluminescence efficiency.

Particularly noteworthy is the potential of lanthanide-based OLEDs to operate in the near-infrared (NIR) region, owing to the inherently low-energy 4f–4f transitions of certain lanthanides (e.g., $Nd^{3+}$, $Yb^{3+}$, $Er^{3+}$). NIR-emissive OLEDs are of growing interest for applications in biosensing, phototherapy, night vision displays, and secure communications. The spectral purity, low self-absorption, and characteristic emission profiles of lanthanides make them highly suitable for this spectral range. However, improving their charge transport and reducing their excited-state lifetimes remain active areas of research critical for their practical deployment.

## Section 2. Plasmonic Nanostructures for Enhancing the Performance of OLEDs

The integration of nanomaterials into OLED architectures opens new pathways for enhancing device performance. Among them, noble metal nanoparticles – especially gold and silver – are of particular interest due to their localized surface plasmon resonance (LSPR), which can amplify emission, improve charge injection, and enhance light outcoupling (**Figure 3**). These effects are especially relevant for lanthanide-based emitters, which suffer from inherently long excited-state lifetimes and low radiative rates.

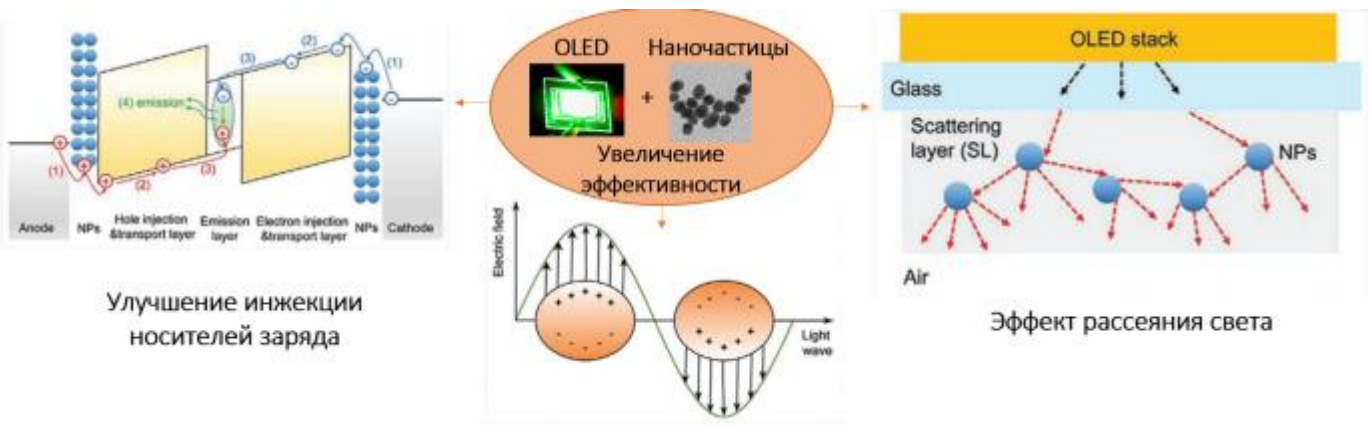

**Figure 3.** The effects that occur when using nanoparticles in OLED.

Plasmonic**s** offers a promising strategy to overcome these limitations by increasing the radiative decay rate and shortening the excited-state lifetime, thereby improving both efficiency and stability of lanthanide-based OLEDs. Gold nanoparticles, in particular, combine high chemical stability, tunable LSPR in the visible and NIR regions, and excellent processability – making them ideal candidates for plasmon-assisted photonic engineering in OLEDs.

### 2.1. Unique Properties of Nanostructures and Their Relevance for OLED Applications

Nanoparticles (NPs) possess a suite of unique physicochemical properties that fundamentally differ from those of their bulk counterparts. These include high mechanical strength, enhanced surface-to-volume ratio, chemical reactivity, quantum confinement, and strong optical resonances – all of which can be tuned via careful control over the NP's size, morphology, composition, and the surrounding dielectric environment. Such tunability is particularly appealing for optoelectronic technologies, including OLEDs, where precise control over emission, charge transport, and energy transfer processes is essential for performance optimization[28–30].

### Optical Properties: Size- and Shape-Dependent Phenomena

The optical behavior of nanoparticles, especially noble metal NPs (e.g., Au, Ag), is largely governed by localized surface plasmon resonance (LSPR). This collective oscillation of conduction electrons at the NP surface, when driven by incident electromagnetic radiation. LSPRs give rise to intense absorption and scattering of the incident light with high electromagnetic fields concentration at NPs' surface. Importantly, the resonance wavelength is not fixed but varies with the NP's geometry, composition, and dielectric surroundings[28,31]. This phenomenon enables subwavelength engineering of optical fields, with strong implications for OLED emission efficiency, spectral tuning, and outcoupling (**Figure 4**)[28].

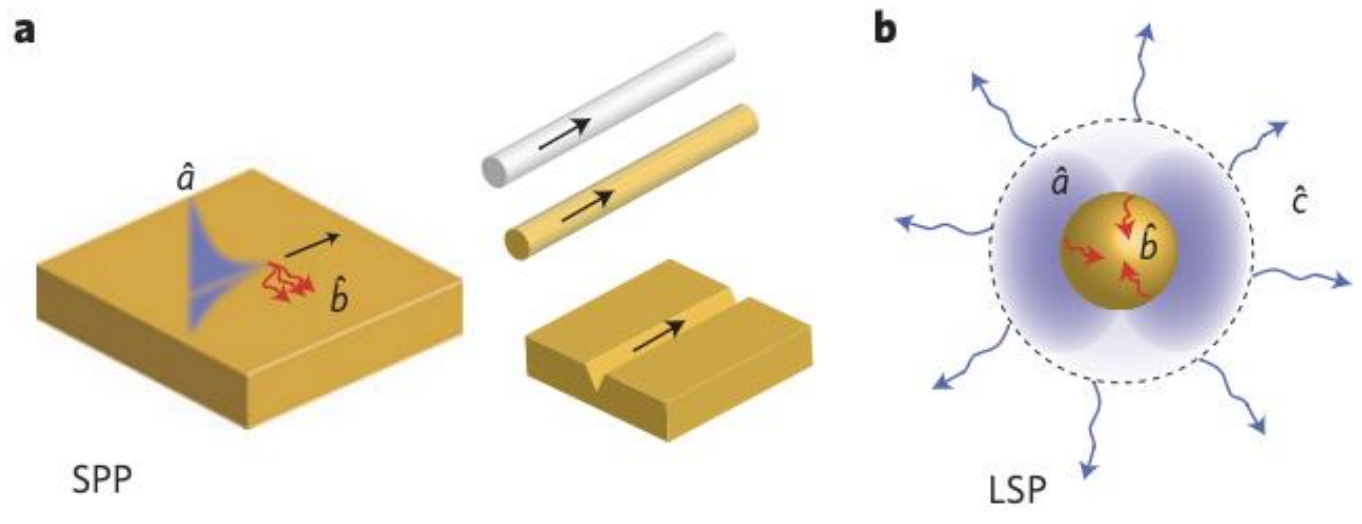

**Figure 4.** Quantization of SPPs on waveguides and LSPs at nanoparticles. a) The magnitude of the transverse z component of the mode function $U_K(r)$ for a single SPP excitation, denoted by aˆ, is shown along with a selection of alternative waveguide geometries. b) The magnitude of the radial r component of the mode function $U_l(r)$ for a single LSP excitation. Reservoir modes, denoted by bˆ and ˆc, are also shown for both SPPs and LSPs, enabling loss to be included in the quantization. Reproduced from [DOI: 10.1038/NPHYS2615].

Recent studies have demonstrated the ability to shift and amplify LSPR by manipulating the aspect ratio of gold nanorods [32], or through core–shell architectures such as Au@$SiO_2$, where plasmon-exciton interactions can be exploited. For instance, by embedding plasmonic NPs in the emissive layer or adjacent transport layers of OLEDs, researchers have achieved enhanced photoluminescence and electroluminescence through near-field coupling mechanisms.

Moreover, semiconductor quantum dots (QDs) represent another class of optically active NPs with size-tunable emission. Due to quantum confinement effects, the bandgap of QDs widens as their size decreases, enabling precise spectral control across the visible and near-infrared ranges. This is particularly relevant for dual-emissive lanthanide systems or NIR-emitting OLEDs, where traditional emitters struggle with low quantum yields and poor overlap with standard charge transport layers.

### Magnetic Properties: Nanoscale Control and Spin-Related Phenomena

Magnetic nanoparticles (MNPs), particularly those based on iron oxides or transition metals, exhibit distinctive properties when confined to the nanoscale. Below the critical size threshold (typically 20-30 nm), many MNPs enter the superparamagnetic regime, characterized by high magnetic susceptibility and zero remanence. These properties can be finely adjusted by tuning the size distribution, crystallinity, and surface chemistry of the nanoparticles[33].

Importantly, MNPs in the form of the nanoclusters have recently emerged as promising tools for modulating excitonic processes in OLEDs. It has been shown that the presence of a magnetic field or magnetic dipole interactions from embedded Co nanoclusters can influence singlet–triplet interconversion rates via spin-mixing mechanisms, leading to enhanced intensity of electroluminescence[34]. This approach represents a novel strategy to exploit the spin degree of freedom in organic emitters without the need for heavy-atom phosphorescent dopants.

Notably, even traditionally non-magnetic materials like Pt can display induced magnetic behavior at the nanoscale, opening additional avenues for tailoring NP functionality beyond their bulk electronic structure[35].

### Size Effects and Multifunctional Integration

Across all NP classes — metallic, magnetic, and semiconductor — size plays a dominant role in dictating properties. For noble metal NPs, LSPR is highly sensitive to particle geometry and size, while in semiconductor QDs, the energy gap directly scales with size due to confinement. In MNPs, coercivity, saturation magnetization, and blocking temperature exhibit non-linear dependencies on particle size. These trends have motivated extensive efforts toward precision synthesis protocols that ensure narrow size distributions, monodispersity, and robust colloidal stability.

The convergence of optical and magnetic functionalities in a single nanoscale platform makes NPs particularly attractive for next-generation OLEDs. For example, composite systems combining LSPR-active metal NPs and magnetic cores have been proposed to simultaneously enhance exciton generation and control spin statistics. Furthermore, the facile integration of NPs into OLED architectures — either as dispersed additives or structured layers — allows for the incorporation of field-enhancing, light-managing, or spin-modulating features without compromising processability.

## 2.2. Localized Surface Plasmon Resonance Effect of Nanoparticles on the Photoluminescence of Lanthanide Complexes

Lanthanide complexes are well known for their sharp emission lines, long excited-state lifetimes, and high color purity, making them attractive candidates for optoelectronic applications, including OLEDs. However, their inherently low molar absorption coefficients and forbidden 4f–4f transitions often limit luminescence efficiency. One promising strategy to overcome these limitations is the use of localized surface plasmon resonance (LSPR) supported by metallic nanoparticles. When properly engineered, the intense near-field generated by LSPR can enhance both the absorption and emission processes of nearby lanthanide emitters, resulting in increased radiative rates and improved photoluminescence quantum yields. Moreover, this approach may enable directional emission, lifetime tuning, and spectral modulation—parameters critical for the development of high-performance photonic devices.

In this context, several studies have demonstrated the capability of plasmonic nanoparticles, particularly silver and gold, to enhance the photoluminescence of various lanthanide complexes. The design and morphology of these nanoparticles—especially their shape, size, and dielectric environment—play a crucial role in dictating the spectral overlap with lanthanide transitions and in achieving optimal emitter–plasmon coupling. Below, we review recent key studies that illustrate the effectiveness of LSPR-active nanoparticles in modulating the luminescence properties of lanthanide-based materials.

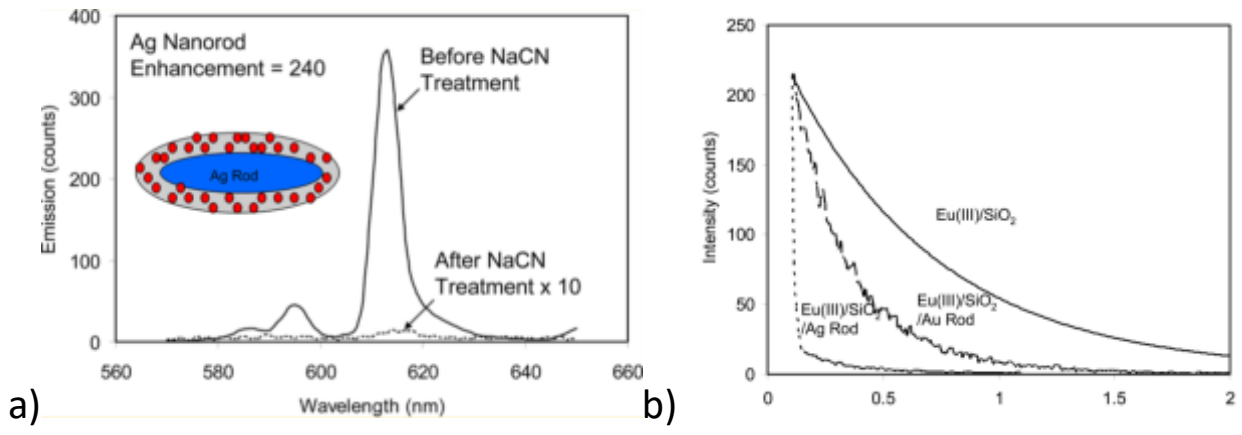

**Figure 5.** a) Emission spectral changes for the Eu (III) chelates on the Ag rods in aqueous solution prior to and after NaCN treatment, b) Decay curves of Eu (III) chelates on the metal-free silica templates, silver rods, and gold rods

In the work of Zhang et al.[51] investigated the influence of the plasmon resonance of silver nanorods on the optical properties of europium (III) chelates was systematically investigated *via* near-field interactions. Silver nanorods have two distinct plasmon absorption bands, one corresponding to longitudinal and the other to transverse plasmon oscillations. The absorption maxima occur at 394 and 675 nm, respectively. These plasmon resonances are strategically aligned with the absorption and emission bands of the chelates, providing optimal spectral overlap for effective plasmon-fluorophore interaction. The europium chelates were immobilised in thin layers of silicon dioxide that were deposited on the surface of the silver nanorods. This allowed the fluorophores to be placed within the near-field zone of the surface plasmons, while maintaining an optimal distance that enhanced the luminescence. The key result of the study was a remarkable 240-fold increase in the emission intensity of Eu (III) chelates compared to control samples in metal-free silicate matrices – setting a new benchmark for plasmon-enhanced lanthanide luminescence (**Figure 5a**). This gain significantly exceeds similar values for europium chelates on gold nanorods or silver nanospheres, highlighting the advantages of using silver nanorods with two plasmon resonances. Plasmon-induced amplification was accompanied by a substantial decrease in the luminescence lifetime of the chelates (from 0.5 ms to 10 µs), indicating a strong increase in the radiative transition rate resulting from near-field interaction with the nanorods' plasmon modes (**Figure 5b**).

Kang et al.[52] investigated the effect of core-shell Ag@$SiO_2$ nanoparticles on the luminescent properties of rare earth element complexes of samarium and dysprosium with various benzoic acid derivatives. Two types of spherical Ag@$SiO_2$ nanoparticles with silicate shell thicknesses of 26 and 41 nm were synthesised using the modified Steber method. Eight types of Sm(III) and Dy(III) complexes were also synthesized. The luminescent properties of lanthanide complexes are significantly enhanced during interaction with Ag@$SiO_2$ nanoparticles due to the localized surface plasmon resonance effect of the silver core. The most pronounced effect was observed for dysprosium complexes with m-aminobenzoic acid: the luminescence intensity increased by a factor of 2.3 when Ag@$SiO_2$ nanoparticles with a shell thickness of 41 nm were used (**Figure 6**).

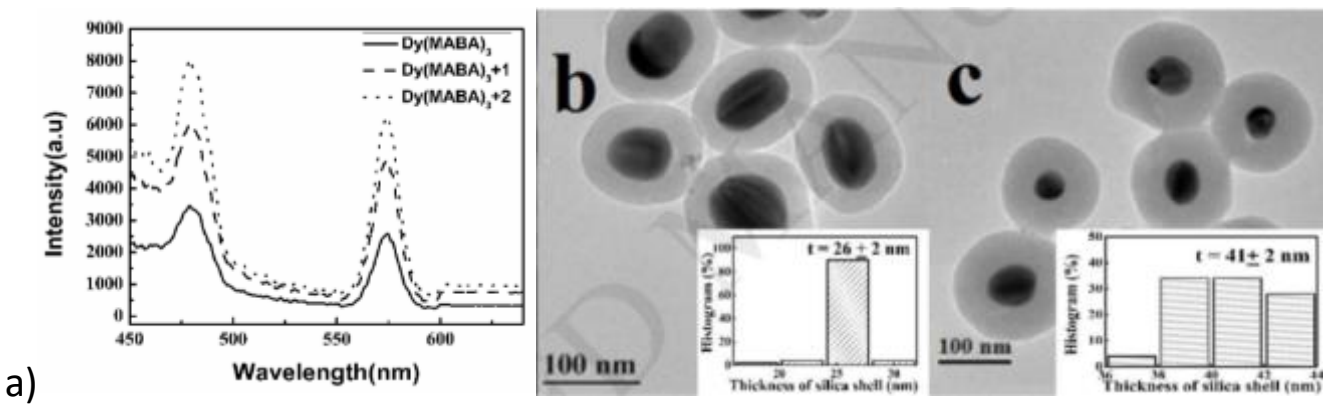

**Figure 6.** a) Fluorescence intensity of $Dy(BA)_3{\cdot}H_2O$ function with Ag@SiO2 nanoparticles, b,c) TEM images of core–shell Ag@$SiO_2$ nanoparticles with the thickness of the $SiO_2$ at 26 and 41nm.

The study of Zhou[53] comprehensively explores the potential for enhancing the fluorescence of terbium complexes by incorporating silver nanoparticles with a silicate shell. To create a metal fluorescence amplification system, monodisperse core-shell Ag@$SiO_2$ nanoparticles with varying silicate coating thicknesses (10, 15 and 25 nm) were developed (**Figure 7a**). The experimental data show a significant increase in the fluorescence of the neodymium complexes when Ag@$SiO_2$ nanoparticles are introduced due to plasmonic effects. The maximum increase in radiation intensity was recorded for systems with a 25 nm thick silicate shell, where the Tb(o-FBA)$_3{\cdot}2H_2O$ complex exhibited a 2.25-fold increase in fluorescence (**Figure 7b**).

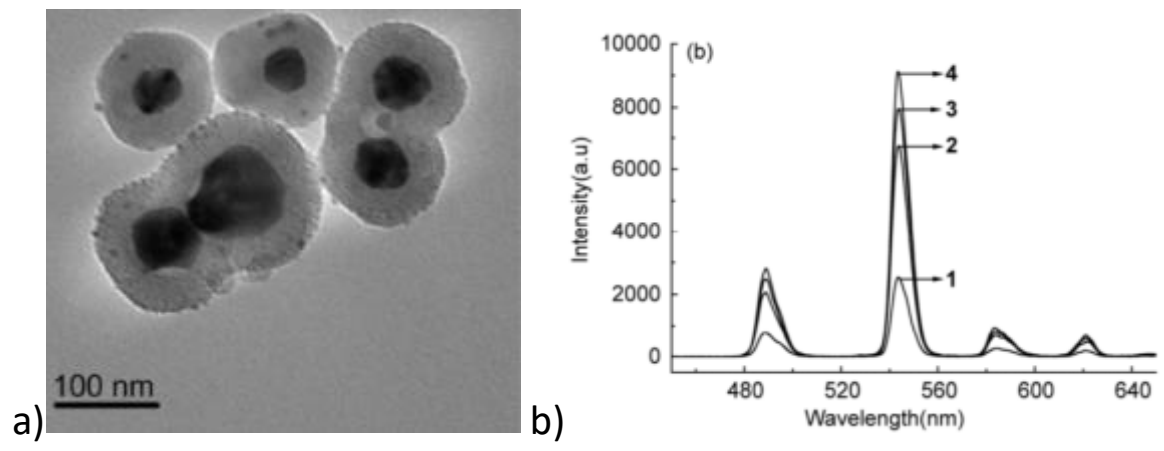

**Figure 7.** a) TEM images of Ag nanoparticles with silica thicknesses of 25 nm, b) the emission spectra of Tb(p-FBA)$_3$ w/o and with Ag@SiO2 nanoparticles A, B and C with silica thicknesses of 10, 15, 25 nm, respectively.

These findings collectively demonstrate the strong potential of plasmonic nanoparticles to enhance the photoluminescence of lanthanide complexes, not only by intensifying radiative transitions but also by reducing emission lifetimes. The precise engineering of shell thickness and spectral overlap enables tunable control of optical properties, paving the way for the development of highly efficient luminescent materials. Applications may range from biomedical imaging and solar energy conversion to advanced OLEDs and photonic circuitry.

### 2.3. Control photoluminescence with Plasmonic Metasurfaces

While individual nanoparticles offer significant photoluminescence enhancement through localized plasmonic fields, recent advances have extended these concepts to ordered plasmonic architectures. In particular, metasurfaces and plasmonic crystals allow for not only amplification, but also angular control and spectral shaping of lanthanide emission. These approaches are discussed in detail in the following section.

One of the most promising strategies to enhance and direct the emission of lanthanide-based complexes involves coupling with surface plasmons supported by plasmonic crystals (PCs)[69–74]. These structures, comprising one- or two-dimensional periodic metal surface or arrays of metal nanoparticles, support propagating surface plasmon polaritons (SPPs) [10.1038/nature01937] or surface lattice resonances [10.1021/acs.chemrev.8b00243], respectively. Their selective interaction with incident light and re-radiation into the far field—governed by the dispersion of plasmonic modes—enables precise control over both the intensity and angular distribution of emission from lanthanide-based emitters.

In a seminal work by Dyakov et al.[69], the impact of plasmonic modes on the photoluminescence (PL) of silicon nanocrystals was systematically studied using one-dimensional arrays of gold stripes with varying widths (30–180 nm) deposited on quartz substrates coated with a thin silica layer embedding the nanocrystals (**Figure 8a**). The plasmonic mode type was shown to depend on the metal filling fraction of the arrays (**Figure 8b**). A key finding was a threefold PL enhancement even with nanoscale air gaps as narrow as 30 nm, corresponding to less than 7% of the structural period (**Figure 8c**). Experimental data exhibited excellent agreement with rigorous coupled-wave analysis (RCWA), revealing a mode transition—from propagating SPPs to quasi-guided modes and localized surface plasmons—as the stripe width increased. This work elucidates the fundamental mechanisms of metal–semiconductor coupling and provides a blueprint for enhancing emission in traditionally low-efficiency systems such as silicon.

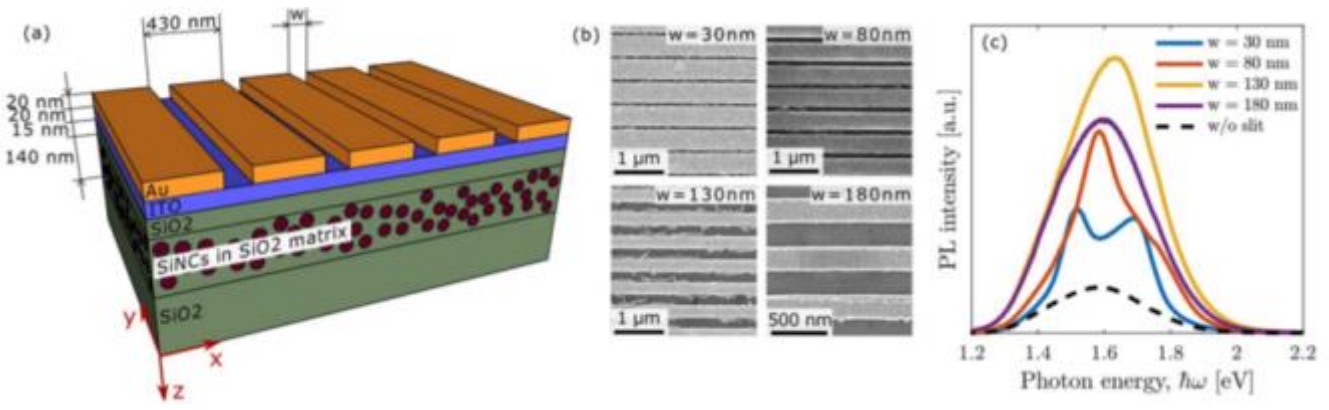

**Figure 8.** a) Schematic of Si nanocrystals (dots) embedded under a gold grating. b) SEM images of the fabricated structures. c) PL spectra for varying air-gap widths; dashed line: control (no gaps).

Dabard et al.[70] addressed the challenge of photoluminescence degradation in two-dimensional mercury telluride (HgTe) nanoplatelets (NPLs), focusing on surface passivation. A multi-step ligand exchange protocol—comprising thiol ligands followed by oleic acid—significantly extended PL stability from one day to two weeks and increased the carrier lifetime from ~1.2 ns to ~4.8 ns. This enabled further integration of HgTe NPLs into hybrid systems with gold plasmonic gratings, leading to pronounced enhancements in both emission directionality and intensity (directionality factor up to 8.5, **Figure 9a**). The final outcome was the fabrication of an infrared light-emitting diode (IR-LED) operating at ~1300 nm, within the telecommunication band, with a turn-on voltage of ~3.5 V (**Figure 9b,c**).

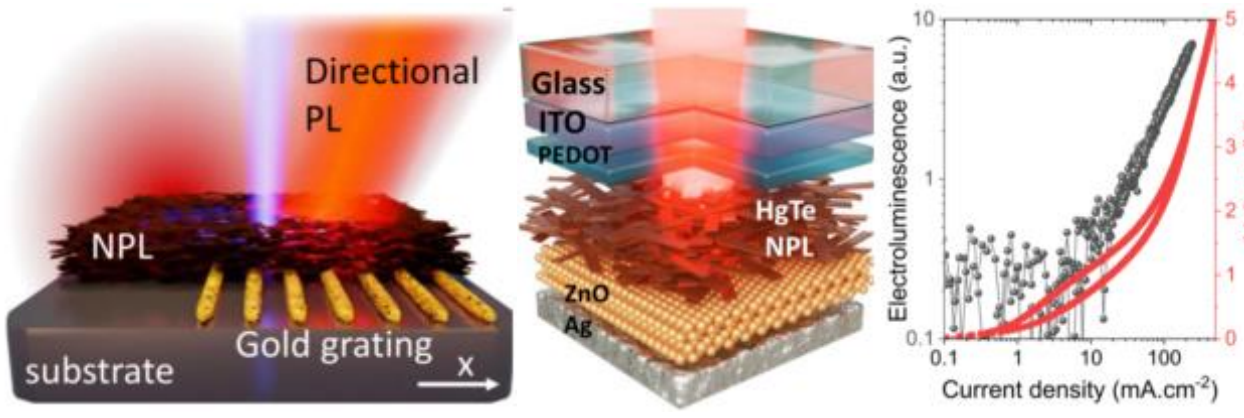

**Figure 9.** Scheme of HgTe NPL film integrated with a gold grating with architecture of IR-LED based on HgTe NPLs and Electroluminescence vs current density and bias voltage.

In a complementary study, Bossavit et al.[74] investigated the size-dependent optical properties of HgTe nanocrystals, revealing strong quantum confinement effects. By tuning the particle diameter to 5–6 nm, the absorption peak could be shifted to 1100 nm, with a corresponding PL maximum redshifted to 1250 nm (**Figure 10**). Time-resolved PL analysis showed a multiexponential decay with fast (~10 ns) and slow (hundreds of ns) components, suggesting a significant contribution from trap-assisted recombination. Transient absorption spectroscopy further confirmed the progressive population of deep trap states, resulting in dynamic redshifts in emission spectra.

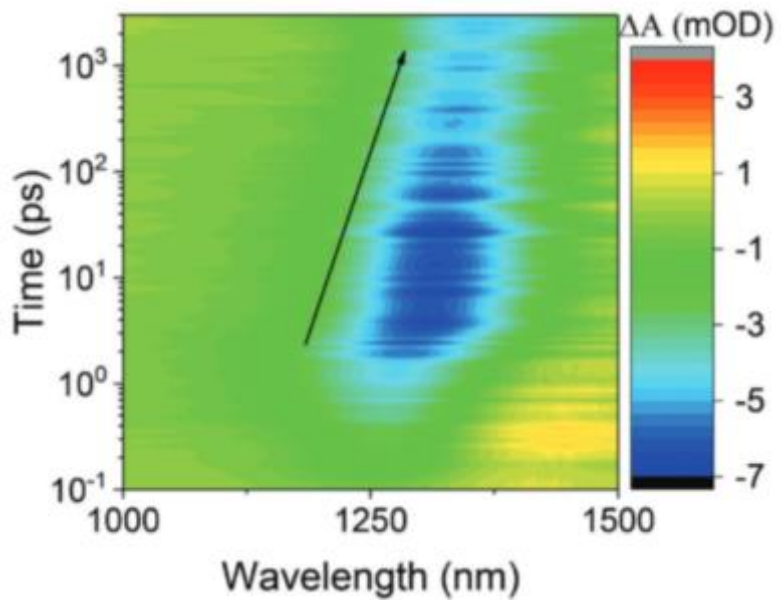

**Figure 10.** 2D transient absorption map of HgTe NPLs in solution: wavelength vs time.

To improve emission characteristics, the authors integrated the NPL films with plasmonic gratings featuring periods of 600–900 nm. Electromagnetic simulations demonstrated strong field confinement at the interface between the gold gratings and HgTe layers. Geometric optimization (300 nm stripe width, 200 nm film thickness) resulted in up to fourfold PL enhancement without altering the spectral shape (**Figure 11**). Directionality arose from resonant coupling between the dipole moments of the nanocrystals and plasmonic lattice modes, highlighting the potential of these hybrid structures as efficient infrared light sources.

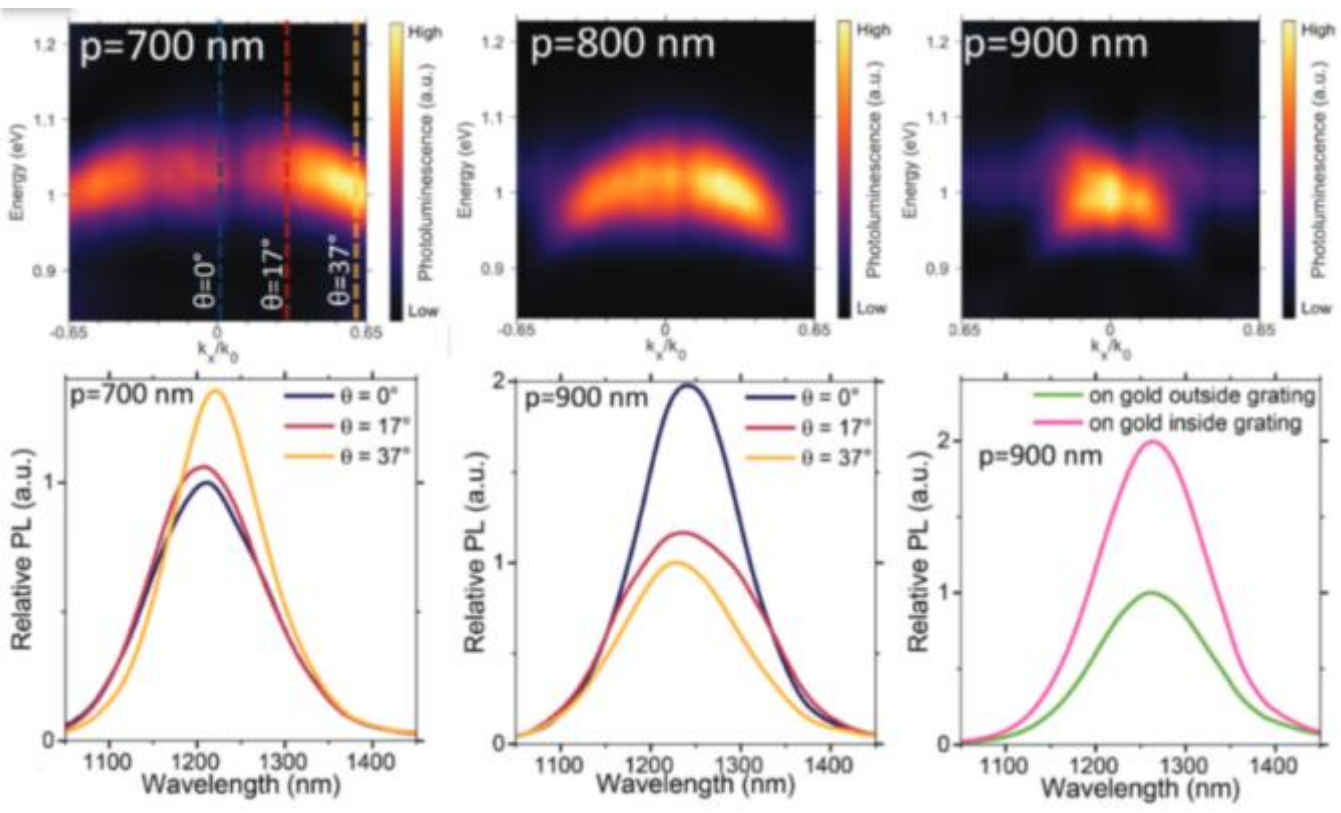

**Figure 11.** a–c) Dispersion maps and PL spectra for plasmonic gratings with periods of 700, 800, and 900 nm.

These findings underscore the dual nature of plasmonic effects in luminescence modulation. While metallic proximity can quench emission via nonradiative decay channels, properly engineered plasmonic architectures can suppress these losses and even amplify the emission through resonance effects. Such hybrid systems open a path toward highly efficient, directionally controlled luminescent devices, which are particularly relevant for narrowband optical sources, sensors, and biomedical diagnostics in the near- and mid-infrared regions.

### 2.2. Enhancement of Charge Carrier Injection and Transport in OLEDs

It is widely acknowledged that in most OLED displays, charge transport is significantly limited by the efficiency of charge carrier injection. This injection is often hindered by high energy barriers at the electrode/organic interface[36], primarily due to the large energy gap between the HOMO–LUMO levels of organic semiconductors and the Fermi levels of the electrodes. The introduction of gold nanoparticles (Au NPs) at the anode/organic and cathode/organic interfaces has been shown to improve charge injection by several mechanisms:

1. lowering the injection barrier through interfacial doping, enabling tunneling;
2. modifying the electrode's work function;
3. enhancing the local interfacial electric field due to electrode surface modification.

Recent studies have demonstrated that the use of nanoparticles is an effective and versatile strategy to improve carrier injection at the electrode/organic interface, owing to their controllable and unique electronic properties. In particular, Au NPs [37–39] have been incorporated into various OLED layers to facilitate enhanced charge injection.

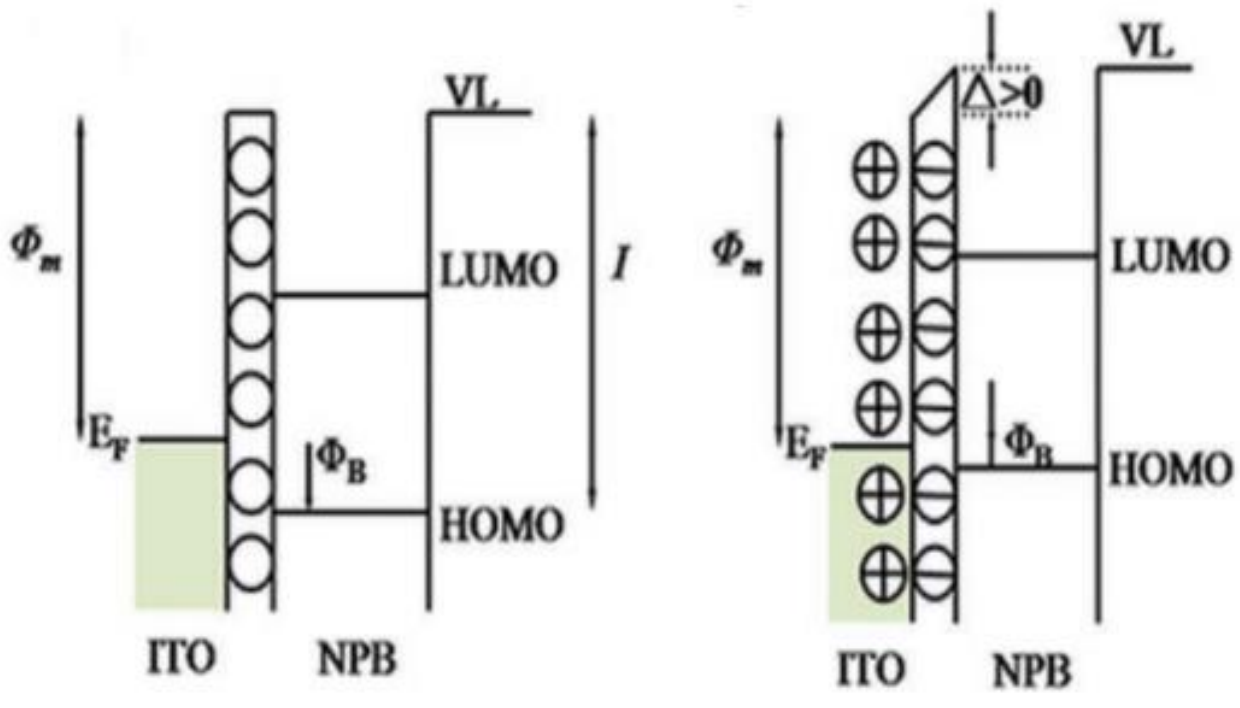

**Figure 12.** Energy-level diagram of the ITO/Au-NPs/NPB interface in the absence (left) and presence (right) of an applied voltage, illustrating dipole effects at the interface.

In the study by Fujita et al.[37], gold nanoparticles stabilized with hyperbranched polystyrene were synthesized and deposited on the ITO surface to improve hole injection in OLED heterostructures (**Figure 12**). This modification led to a significant reduction in turn-on voltage, along with a sixfold increase in current density and brightness. These improvements were attributed to the reduction of the injection barrier via local field effects generated by negatively charged Au NPs.

Similarly, Jeganathan et al.[38] inserted gold nanoparticles between the hole injection layer (HIL) and the anode, resulting in a decrease in turn-on voltage from 7.0 V to 3.0 V and an increase in maximum luminance from 4500 cd/m² to 9000 cd/m². The authors attributed these improvements to energy level alignment at the interface, which reduces the barrier for carrier injection.

Zhang et al.[39] reported the development of a modified hole injection layer composed of $MoO_x$ doped with Au NPs. The use of this mixed layer significantly reduced the hole injection barrier and contact resistance (**Figure 13**), leading to nearly a threefold enhancement in current efficiency.

More intriguingly, Wang et al.[40] introduced nanoscale charge injection "hot spots" on the ITO surface to amplify the local electric field and thereby promote efficient carrier injection (**Figure 14**). In this approach, small-sized aluminum nanoparticles were deposited via magnetron sputtering onto the ITO substrate to act as injection hot spots. In addition to Al NPs, $C_{60}$ fullerenes and its derivative, phenyl-$C_{61}$-butyric acid methyl ester (PCBM), were also shown to serve as effective electron injection hot spots [41].

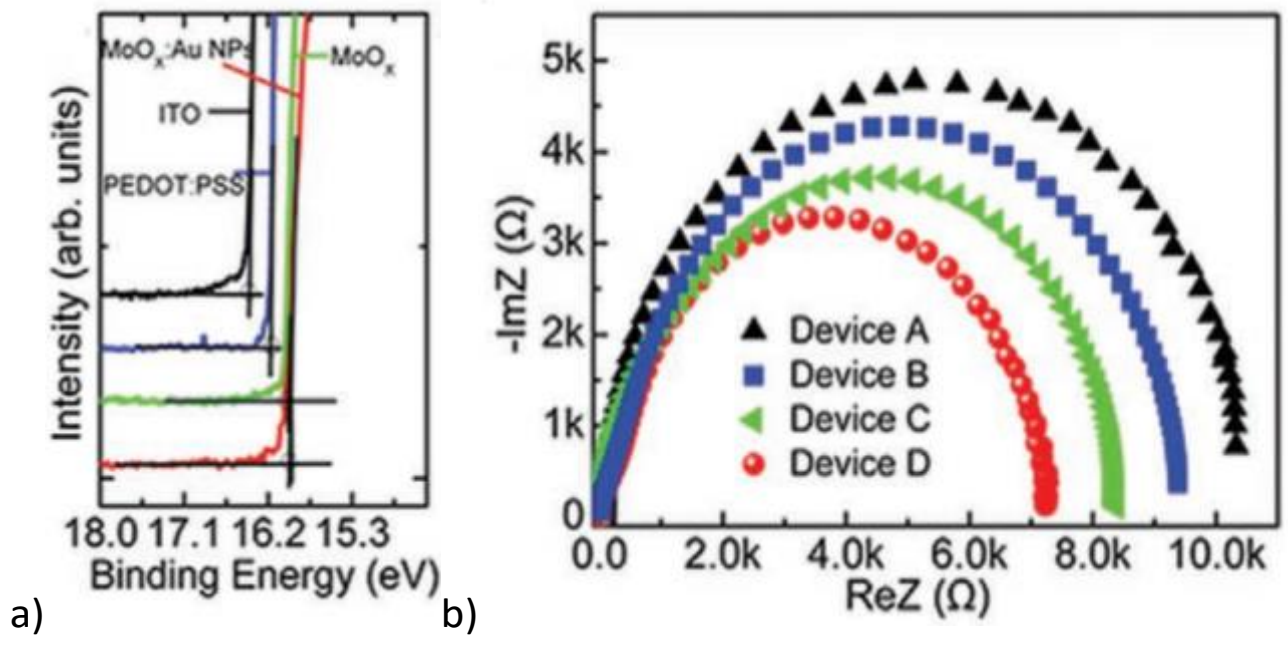

**Figure 13.** a) Ultraviolet photoelectron spectroscopy (UPS) of ITO, ITO/PEDOT:PSS, ITO/$MoO_x$, and ITO/Au NPs:$MoO_x$; b) Impedance spectra of OLED devices without HTL, and with PEDOT:PSS, $MoO_x$, and Au NPs:$MoO_x$, respectively.

The use of such hot spots significantly reduced turn-on voltages and improved device performance in both inverted fluorescent and phosphorescent OLED architectures. Notably, this enhancement was achieved without altering the electrode's work function or optical transparency, as the local field effect is induced solely by the conductive nanostructures. These localized electric fields facilitate electron injection into the conductive nanodomains, providing an alternative strategy for improving injection efficiency beyond conventional barrier engineering.

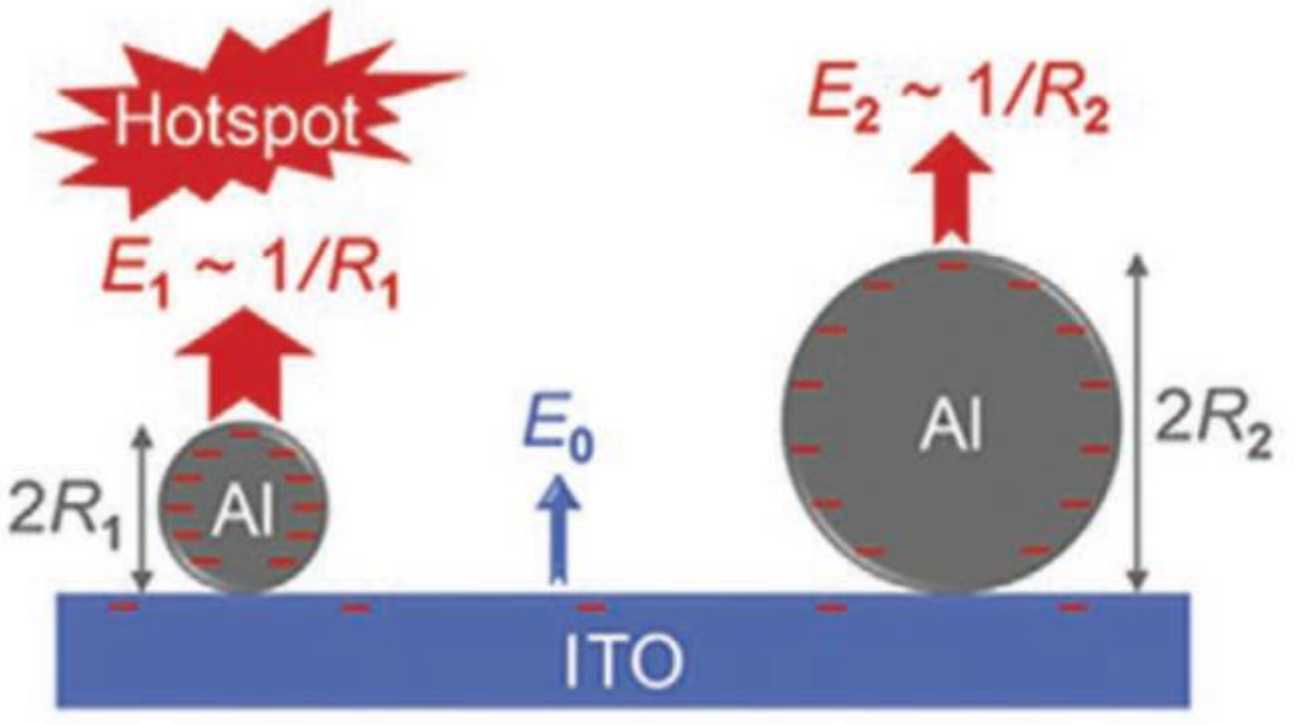

**Figure 14.** Schematic illustration of the effects of electron injection, which shows large and small Al nanoparticles on the ITO surface for comparison.

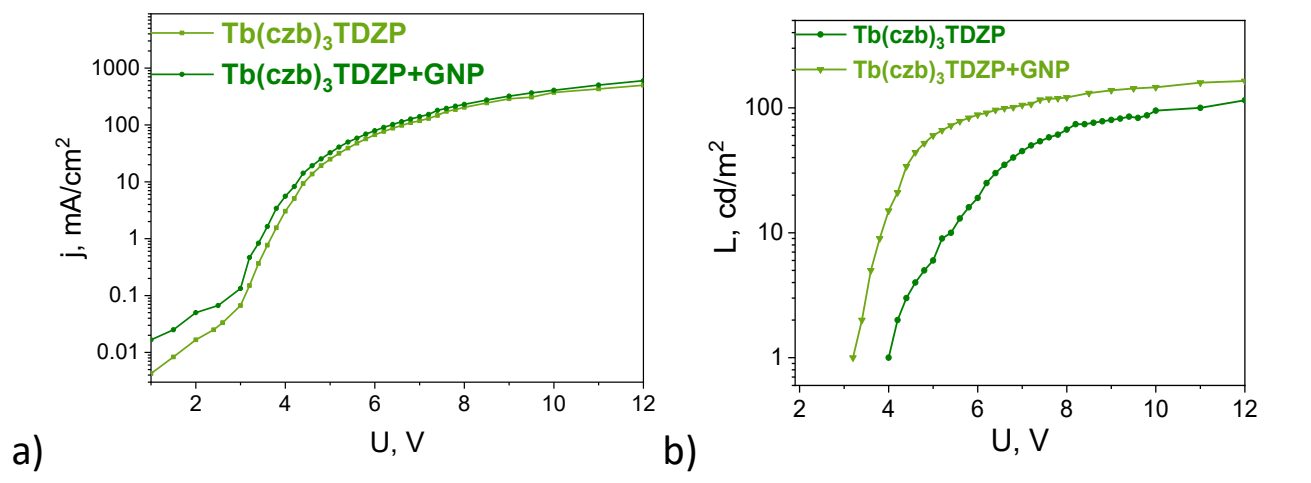

**Figure 15.** a) J-V and b) L-V curves of OLED based on terbium complexes.

Recently, our research group[42] tested OLEDs based on terbium complexes, where nanoparticles were doped into a hole-injection layer PEDOT:PSS. It was found that OLEDs doped with gold nanoparticles have a similar J-V behaviour compared to OLEDs without gold nanoparticles (**Figure 15a**). The slightly higher current density of OLED displays doped with GNP may be due to improved injection. This is confirmed by a decrease in the switching voltage from 4 V for unalloyed OLEDs to 3.2 V for OLEDs doped with GNP, as the energy barrier at the electrode/organic layer interface is reduced (**Figure 15b**).

Thus, the use of gold nanoparticles makes it possible to obtain OLED devices with significantly low switching voltages. The energy structure of lanthanide coordination complexes makes it possible to achieve low operating voltages (3-5 V) due to optimal alignment of energy levels with conventional charge transport materials such as NPB and Alq3. The wide band gap (3.1-4.0 eV) ensures effective carrier limitation in the active layer, reducing leakage currents and increasing the overall energy efficiency of devices. Such indicators are especially important for wearable biomedical electronics, where energy consumption reduction is critically important.

### 2.3. Enhancement of Light Extraction Efficiency in OLED Devices

Light scattering refers to the redirection of electromagnetic waves through the deliberate introduction of scattering layers into the emitted light path, aiming to enhance light extraction efficiency. Nanoparticles are ideally suited for this purpose due to their excellent light-scattering capabilities.

The effectiveness of scattering depends strongly on the size of the nanoparticles, which should be comparable to the wavelength of the emitted light. While smaller nanoparticles—significantly smaller than the emission wavelength—exhibit limited scattering efficiency, they can be employed to increase the effective refractive index.

There are two general strategies for integrating nanoparticles to enhance light extraction:

1. deposition outside the functional layer[43];
2. incorporation directly into the OLED heterostructure[44].

Placing nanoparticles away from the functional layer (typically on the rear side of the glass substrate), known as the external scattering approach, is considered an optimal method for improving outcoupling efficiency. In addition to its high light extraction performance, this method offers several advantages:

1. nanoparticles are located far from the active layers, minimizing their impact on OLED operation;
2. angular color shift can be reduced due to the stochastic nature of the scattering process;
3. OLEDs can be fabricated on a large scale using simple and cost-effective techniques.

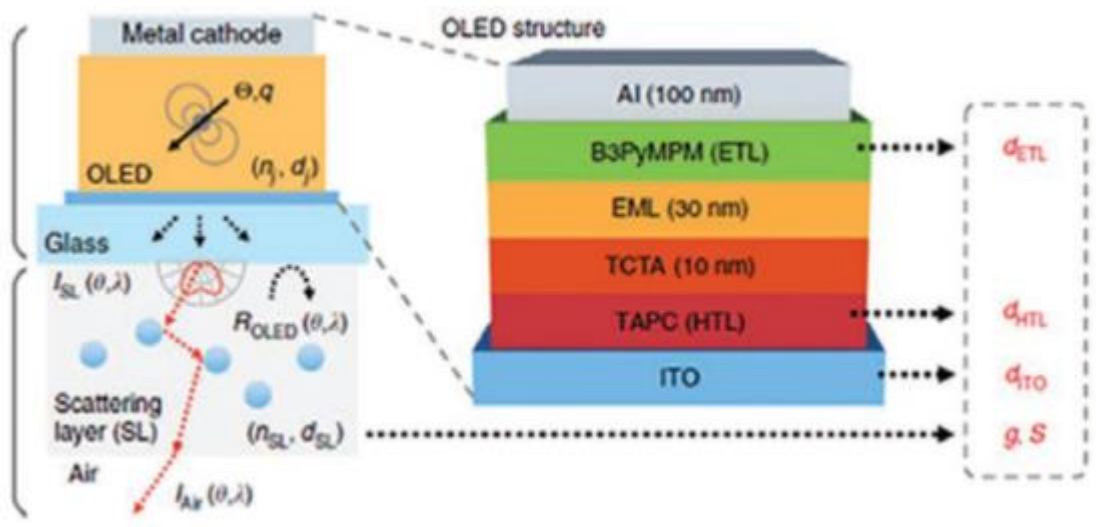

**Figure 16.** Schematic of the proposed 3D optical simulation model, OLED device structure, and parameters used for multidimensional analysis.

Recently, the group of Yoo[45] proposed a highly efficient OLED architecture by introducing an external scattering layer composed of $SiO_2$/polymer nanoparticles on the back side of the glass substrate (**Figure 16**). Their global and multiparametric optimization approach demonstrated that precisely engineered scattering can maximize light extraction. The resulting device exhibited a high luminous efficacy of 221.1 lm·W$^{-1}$ and an equivalent external quantum efficiency (EQE) of 56%.

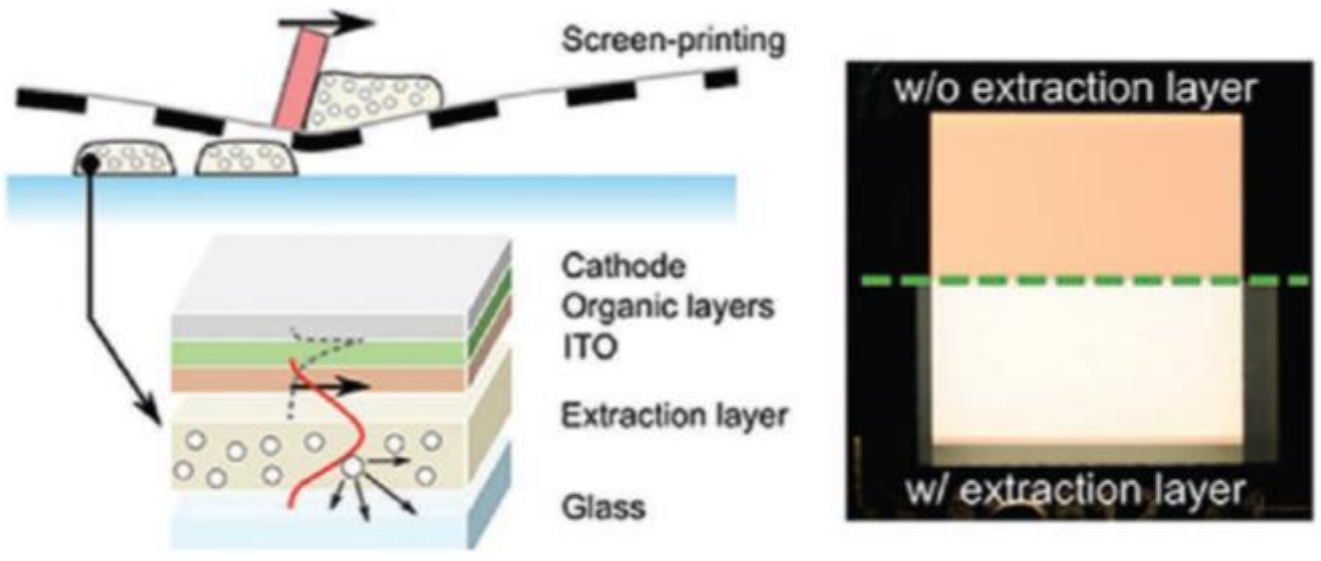

**Figure 17.** Schematic of the stencil-printed internal light-extraction layer between the glass substrate and the OLED stack, and photographs of white OLEDs without and with the internal scattering layer.

Lemmer's group reported a cost-effective stencil printing technique for applying high-refractive-index polymers. A nanocomposite scattering layer of $TiO_2$ nanoparticles was deposited between the glass and the ITO electrode (**Figure 17**, [46]). A 15 × 15 cm² glass substrate was uniformly coated with a 10 µm thick internal extraction layer, demonstrating strong optical enhancement across the visible spectrum. The printed scattering layer yielded a 56% increase in current efficiency, attributed to the effective scattering of emitted light. Furthermore, the angular dependence of emission was significantly reduced. A more stable and blackbody-like emission spectrum was also achieved, with a correlated color temperature of 3065 K and a color rendering index (CRI) of 70.

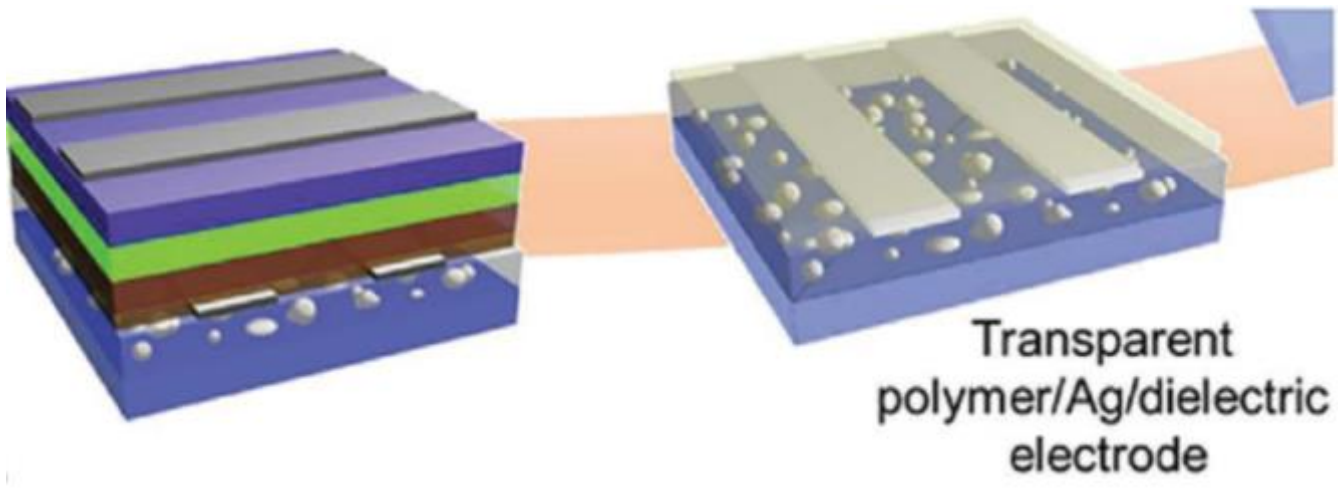

**Figure 18.** Schematic structure of the OLED device incorporating Ag nanoparticles beneath the DMD electrode.

Choi et al.[44] proposed a transparent anode composed of silver nanowires (Ag NWs) integrated with light-scattering $TiO_2$ nanoparticles. Devices employing mesoporous $TiO_2$ NPs exhibited a 30% increase in efficiency compared to those with amorphous $TiO_2$ NPs. This improvement was attributed to stronger light scattering within the OLED heterostructure, driven by higher refractive index contrast and larger surface area of mesoporous NPs. Furthermore, silver nanoparticles were embedded beneath the dielectric/metal/dielectric (DMD) transparent electrode structure to combine the scattering effects of Ag NPs with those of the DMD stack (**Figure 18**, [47]). Ag NPs were formed via rapid thermal annealing of an ultrathin Ag layer—a simple and scalable process. Both experimental and simulation results indicated that such thermal treatment produced silver nanoparticles capable of broadband light scattering, enabling effective outcoupling of waveguided photons. The device incorporating Ag NPs achieved a 25.1% enhancement in efficiency, which was 11% higher than that of the control device without gold nanoparticles.

Direct incorporation of gold nanoparticles into the functional layers of OLEDs results not only in improved light scattering, but also enhances charge injection, as discussed previously. Additionally, noble metal nanoparticles exhibit localized surface plasmon resonance (LSPR) effects, which will be addressed in the following section.

### 2.4. Magnetic Field Effect

Magnetic field effects (MFEs), induced by magnetic materials, provide an effective means of modulating the singlet-to-triplet ratio in fluorescent emitters, offering an alternative strategy for enhancing the efficiency of fluorescent OLED displays. This enhancement arises because MFEs can influence either the hyperfine interaction that governs singlet–triplet interconversion or spin injection from magnetic materials, ultimately increasing the singlet-to-triplet exciton ratio. As a result of magnetic field influence, the singlet fraction can be raised from 1/4 to 1/3, corresponding to a 33% increase.

Hu et al.[48] reported a performance enhancement in single-layer OLED devices through the incorporation of CoFe nanoparticles into a 2-methoxy-5-(2'-ethylhexyloxy)-1,4-phenylenevinylene (MEH-PPV) emissive layer (EML) (**Figure 19a**). The CoFe nanoparticles were fabricated on MEH-PPV/glass substrates, and their ferromagnetic properties were confirmed via magnetic hysteresis loop measurements (**Figure 19b, c**). Incorporation of the nanoparticles into the EML led to an increase in electroluminescence (EL) intensity by 27% in the absence of an external magnetic field and by 32% in its presence. These improvements were attributed to enhanced electron trapping, which increased the exciton formation probability and improved the singlet ratio under the magnetic field (**Figure 19d**).

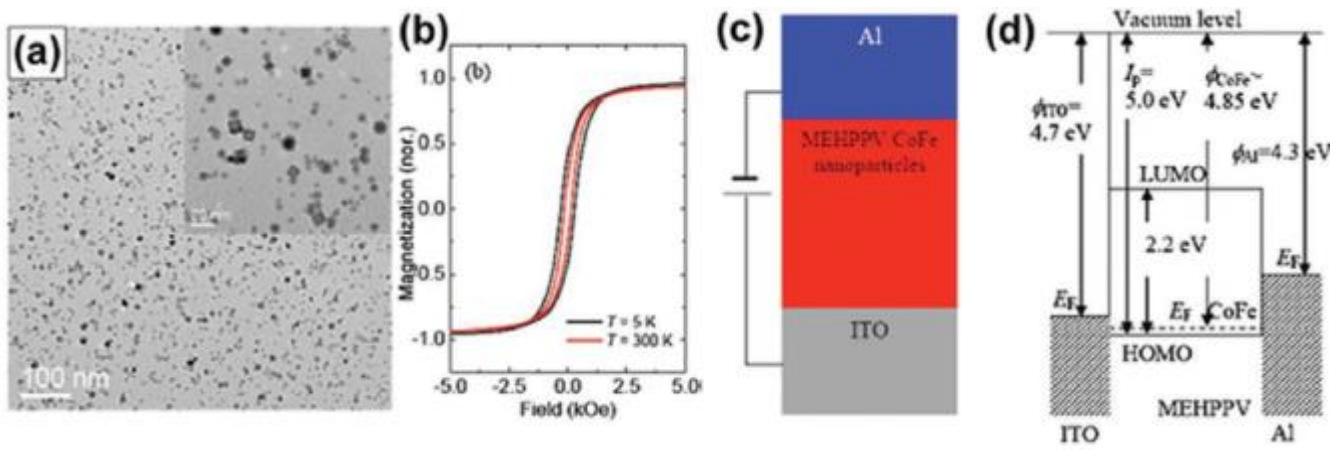

**Figure 19.** (a) TEM images of the CoFe nanoparticles. The inset shows magnified images indicating partial oxidation. (b) Magnetic hysteresis loops of CoFe nanoparticles at 5 K and 300 K. (c) Schematic structure of the single-layer OLED with CoFe-doped EML. (d) Conceptual diagram of the device operation.

Magnetic NPs/PEDOT-based composite hole injection layers (HILs) for efficient fluorescent OLEDs were demonstrated by the group of Zhu[49]. Three types of $Fe_3O_4$-based nanoparticles were compared and optimized: $Fe_3O_4$@$SiO_2$, $Fe_3O_4$@Au, and $Fe_3O_4$@graphene. The current efficiencies of the corresponding devices were significantly improved to 5.13, 5.13, and 4.83

cd·A$^{-1}$, respectively, yielding efficiency enhancements of 35%, 35%, and 38.2% compared to the control device. The observed performance improvement was attributed to the synergistic contributions from light scattering, localized surface plasmon resonance (LSPR), and magnetic field effects (MFEs) associated with the magnetic nanoparticles (MNPs) in the HIL.

Furthermore, Zhang and co-workers investigated the distinct roles of $Fe_3O_4$@Au magnetic nanoparticles in both fluorescent and phosphorescent OLEDs[50]. They demonstrated that LSPR and scattering effects contributed to performance enhancement in both types of devices. However, magnetic field effects were significant only in fluorescent OLEDs, whereas their influence was negligible in phosphorescent OLEDs.

## 2.5. Localized Surface Plasmon Resonance Effect of Nanoparticles for OLED

One of the most important features of metallic nanoparticles is the localized surface plasmon resonance (LSPR) effect. LSPR refers to the coherent oscillations of delocalized conduction electrons, excited by the electromagnetic field of incident light at the metal–dielectric interface. This phenomenon results in the amplification of the electromagnetic field, enhancing the intensity of emission at specific wavelengths[29,31,54].

Of particular relevance to OLEDs is the use of LSPR to increase the radiative rate constant via the Purcell effect, thereby reducing the excited-state lifetime—a crucial strategy for overcoming the intrinsically low energy efficiency of OLEDs based on lanthanide complexes. In optimized configurations with a single emitter, the Purcell factor can reach values as high as $10^3$ [55], although more realistic plasmonic structures typically yield values in the 10–100 range[56–59].

Unlike other photonic applications, enhancement of lanthanide-based coordination compound OLEDs via the Purcell effect does not require an increase in photoluminescence quantum yield (PLQY), but rather a reduction in excited-state lifetimes. In fact, improving PLQY may be counterproductive in this context, as it often correlates with increased competition from nonradiative processes. Lanthanide complexes already exhibit exceptionally high PLQYs—up to 100% for $Tb^{3+}$ [60–62] and around 90% for $Eu^{3+}$ [63]—yet their long luminescence lifetimes, often on the millisecond scale, remain a key limitation.

Numerous studies have explored the integration of gold nanoparticles (GNPs) into OLED architectures. For non-lanthanide OLEDs, the versatility of GNPs has been exploited in various heterostructures[64,65]. In [64], gold nanoparticles with diameters up to 90 nm were incorporated, with $Alq_3$ serving as the emitter. Devices with larger particles (50 and 90 nm) demonstrated increased current density and approximately 20% enhancement in electroluminescence efficiency (**Figure 20**). The authors attributed this improvement to a noticeable reduction in exciton lifetime due to LSPR.

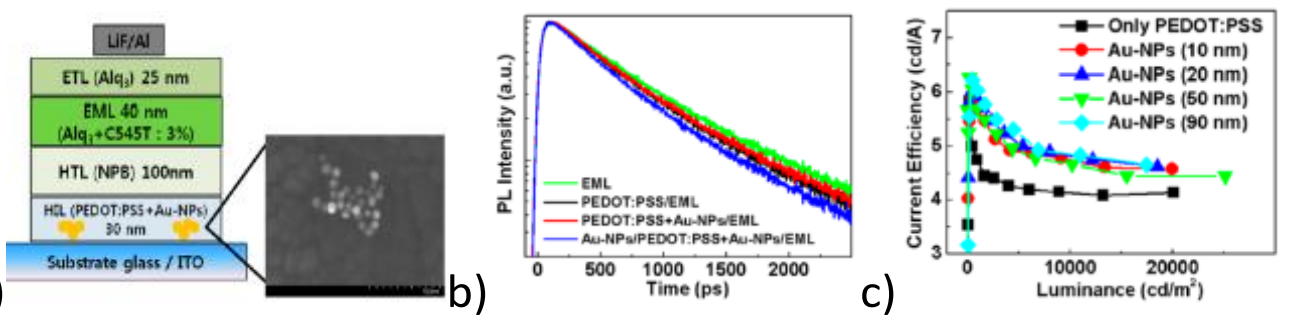

**Figure 20.** (a) OLED heterostructure; (b) measured excited-state lifetime; (c) current efficiency of the OLED.

In another study 65, using C545T as the emitter (**Figure 21**), introducing GNPs into the PEDOT:PSS layer led to a 25% increase in brightness and EL efficiency. Time-resolved photoluminescence spectroscopy confirmed that the enhancement was due to an increased spontaneous emission rate induced by the plasmonic resonance of GNPs.

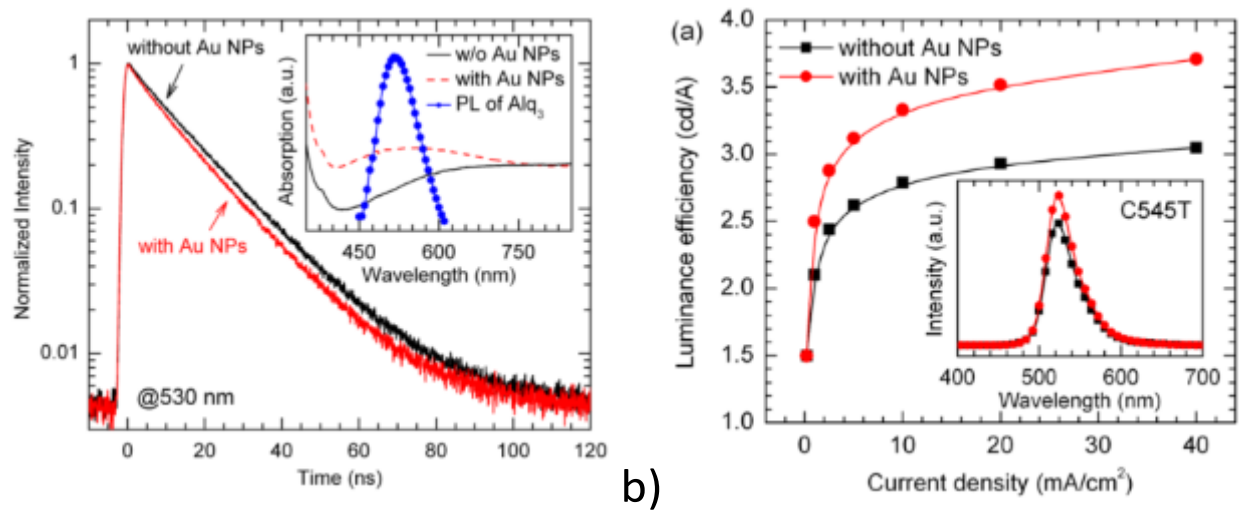

**Figure 21.** (a) Excited-state lifetime; (b) brightness efficiency as a function of current density.

Further work[66] demonstrated that incorporation of GNPs into the PEDOT:PSS hole injection layer led to a substantial increase in power efficiency of OLEDs using $Ir(ppy)_3$ as the emitter—from 29.0 to 36.2 lm/W (a 24.8% increase), along with a maximum brightness increase from 21,550 to 27,810 cd/m² (a 29.1% gain). Notably, a blue OLED based on Flrpic exhibited an efficiency boost of 50% and 35% at luminances of 100 and 1000 cd/m², respectively, under the same structural conditions. This remarkable improvement was attributed to strong short-wavelength light absorption by GNPs, which initiated pronounced LSPR and resulted in significantly enhanced device performance.

GNPs can also facilitate photon extraction from waveguide or substrate modes. Furthermore, due to the near-field coupling range being less than 50 nm, placing the nanoparticles in close proximity to the emissive center increases the local recombination rate[67].

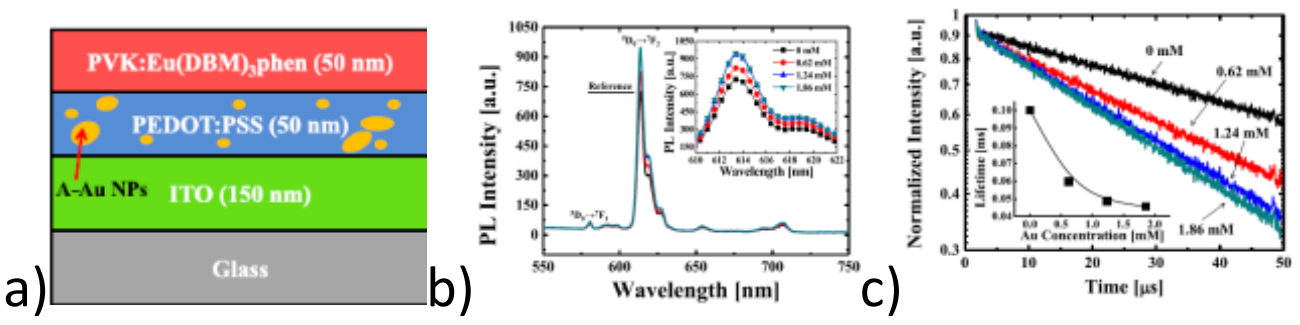

**Figure 22.** (a) OLED heterostructure; (b) photoluminescence spectra; (c) excited-state lifetime.

To date, no extensive studies have explored the LSPR effect of gold nanoparticles in OLEDs based on lanthanide emitters. The only report[68] demonstrated a 64% reduction in the excited-state lifetime of a europium coordination compound, resulting in a 31% enhancement in the quantum efficiency of electroluminescence (**Figure 22**, [68]).

These findings highlight the promising potential of gold plasmonic nanoparticles to address the critical challenge of long lifetimes in lanthanide-based OLEDs, offering a viable pathway toward significantly enhanced energy efficiency in these systems.

## Outlook and Future Directions

The integration of lanthanide-based coordination compounds into organic light-emitting diode (OLED) architectures offers a unique path toward devices with narrowband emission, high photostability, and potential for multifunctional sensing capabilities. However, the intrinsic characteristics of lanthanide ions—such as low oscillator strengths, long lifetimes, and parity-forbidden 4f–4f transitions—pose several challenges when these complexes are embedded into solid-state emissive layers or hybrid metastructures.

**Radiative Inefficiency and Emission Enhancement Strategies**

One of the principal limitations in lanthanide OLEDs arises from the inherently low radiative rates (typically 10–1000 $s^{-1}$) of 4f transitions, which are highly shielded and only weakly coupled to the host matrix. While this contributes to the long excited-state lifetimes and thermal robustness of the emission, it also results in poor electroluminescence efficiency, especially under the current densities required for device operation.

To address this, recent studies—including those reviewed here—highlight the role of plasmonic and dielectric metasurfaces in enhancing lanthanide emission through local field effects and Purcell enhancement. Coupling $Eu^{3+}$ or $Yb^{3+}$ emitters to localized surface plasmon resonances (LSPRs) or lattice modes of gold gratings has been shown to increase both emission intensity and directionality. However, achieving effective coupling requires precise spatial positioning of the emitters relative to the electromagnetic hotspots, which remains a synthetic and fabrication challenge, especially at sub-10 nm scales.

Furthermore, the spectral mismatch between plasmonic resonances and lanthanide emission peaks (often in the visible or near-infrared) limits the effectiveness of metal-enhanced luminescence, especially for deep-red and NIR-emitting ions. This mismatch can be partially resolved by engineering the plasmonic modes via geometry tuning or by using hybrid metal–dielectric metasurfaces with adjustable photonic dispersion.

**Exciton Management and Energy Transfer Efficiency**

Lanthanide complexes typically rely on ligand-to-metal charge transfer (LMCT) or antenna-based sensitization mechanisms to generate luminescence, as direct excitation of 4f states is inefficient. In OLED architectures, this implies a multi-step energy transfer process: charge injection into the host, exciton formation, transfer to the antenna ligand, and finally intramolecular sensitization of the lanthanide ion.

This multistep pathway increases the likelihood of exciton quenching, especially at interfaces or in the presence of non-radiative trap states. The integration of nanostructured lanthanide emitters into complex device stacks therefore requires rigorous surface passivation and ligand engineering to suppress non-radiative channels and to stabilize the emitting states under electrical bias.

The case of HgTe nanoplatelets demonstrates that even in covalently capped nanocrystals, degradation via trap-assisted recombination can drastically reduce device lifetimes unless properly mitigated. For lanthanide complexes, which are even more sensitive to surface quenching, the use of rigid, π-conjugated antenna ligands and sterically protected coordination spheres is essential. Additionally, ligand exchange strategies must be compatible with both charge transport and emissive layer processing conditions.

**Integration with Plasmonic and Photonic Architectures**

As shown in recent work, integrating lanthanide emitters into metasurface-enhanced architectures provides a promising route for directional emission and modulation of photophysical properties. Nevertheless, unlike quantum dots or conjugated polymers, lanthanide emitters are highly anisotropic in both transition dipole orientation and spatial response. This necessitates co-optimization of molecular design and nanophotonic environment to achieve consistent enhancement.

For instance, the placement of $Eu^{3+}$ or $Yb^{3+}$ complexes within high-field regions of a plasmonic lattice must take into account not only the spatial field profile but also the spectral alignment and the orientation of emitting dipoles. Moreover, recent advances in dual-emissive complexes and thermosensitive lanthanide systems introduce an additional design variable: controlling the LIR (luminescence intensity ratio) or lifetime readout in real time, which can be perturbed by cavity coupling and electromagnetic confinement.

Here, metasurfaces can act not only as enhancers but also as modulators of the emission ratio, enabling dynamic tuning of the color temperature or integrating optical thermometry directly into the OLED pixel. This functionality could be particularly valuable in bio-integrated or wearable devices, where real-time monitoring of temperature or physiological parameters is required.

**Charge Transport and Device Engineering Considerations**

A critical, yet often overlooked, limitation of lanthanide complexes in OLEDs is their poor charge transport properties. Unlike small-molecule fluorophores or thermally activated delayed fluorescence (TADF) emitters, lanthanide emitters exhibit localized frontier orbitals and low carrier mobilities. As a result, exciton formation must occur near the emissive site, necessitating fine-tuned device architectures with optimized carrier balance and recombination zones.

Approaches such as co-doping with high-mobility hosts, use of bipolar transport layers, and energy level engineering have shown partial success. However, for nanoparticle-based emitters, charge injection becomes even more challenging due to the presence of insulating ligand shells and potential interfacial barriers. Advanced strategies such as ligand stripping, conductive polymer encapsulation, or in situ cross-linking may provide solutions, albeit at the cost of synthetic complexity.

Thus, the convergence of lanthanide photophysics with metasurface engineering, solution-processable nanoparticle synthesis, and advanced device architectures opens new avenues for OLEDs beyond display technologies—especially in sensing, secure communications, and low-power photonic circuits.

Further developments will likely involve:

- Engineering ligand shells for charge injection compatibility;
- Spectral matching of antenna ligands with optical resonances;
- Multiplexed emission control via photonic band structure engineering;
- Integration with time-resolved or multi-modal readout (e.g., PLIM, LIR, lifetime-based sensing).

In summary, while lanthanide-based nanoparticles present a distinct set of challenges in OLED design, their unique optical signatures and capacity for structural and functional tuning make them a highly attractive class of emitters for next-generation optoelectronic platforms.

## Author Contributions

All authors contributed equally.

## Conflicts of interest

There are no conflicts to declare.

## Acknowledgments

The authors acknowledged support from the Lomonosov Moscow State University Program of Development for providing access to OLED production equipment. This research was performed according to the Development Program of the Interdisciplinary Scientific and Educational School of Lomonosov Moscow State University “Photonic and Quantum Technologies. Digital Medicine.” The study was conducted under the state assignment of Lomonosov Moscow State University. A.Yu. Frolov thanks the Russian Science Foundation (grant no. 24-72-00042).

## Notes and reference